\documentclass[aps,prl,preprint,superscriptaddress,showpacs]{revtex4}
\usepackage{graphicx}
\usepackage{mathrsfs}

\begin{document}

\title{Cavity cooling of a nanomechanical resonator by light scattering}

\author{I. Favero}
\email{ivan.favero@physik.uni-muenchen.de}

\author{K. Karrai}
\email{karrai@lmu.de}

\affiliation{Center for Nanoscience and Fakult\"{a}t f\"{u}r Physik,
Ludwig-Maximilians-Universit\"{a}t, Geschwister-Scholl-Platz 1,
80539 M\"{u}nchen, Germany\\}

\date{13 0ctober 2007}

\begin{abstract}

We present a novel method for opto-mechanical cooling of
sub-wavelength sized nanomechanical resonators. Our scheme uses a
high finesse Fabry-Perot cavity of small mode volume, within which
the nanoresonator is acting as a position-dependant perturbation by
scattering. In return, the back-action induced by the cavity affects
the nanoresonator dynamics and can cool its fluctuations. We
investigate such cavity cooling by scattering for a nanorod
structure and predict that ground-state cooling is within reach.

\end{abstract}


\maketitle

Quantum mechanics describes the behavior of matter on different
length scales from quarks to collective macroscopic states referring
to supra-conductivity and superfluidity. Still, in order to clarify
its transition between microscopic and macroscopic range, several
experimental programs aim at observing quantum phenomena at larger
scales \cite{Legett,Haroche,Bouwmeester}. To this respect, reaching
experimentally the quantum ground state of a macroscopic mechanical
resonator is appealing, as it would at the same time allow to study
in a quantum regime a system with a macroscopic mass, the paradigm
of gravitational interactions \cite{Penrose}.

Experiments using state of the art cryogeny and capacitive detection
techniques have already approached closely the quantum regime for
mechanical oscillators with eigenfrequencies ranging from $10$ MHz
to 1 GHz range \cite{Schwab,Cleland}. The recent development of
optical cooling techniques, either active
\cite{Mancini,Cohadon,BouwmeesterActive} or passive
\cite{ConstanzeNature,Arcizet,Gigan,Kippenberg,IvanAPL,Mavalvala},
creates hope for reaching this regime beyond the possibilities
offered by nowadays cryogenic methods. In optical cooling, the low
noise photons of a laser source are used to extract thermal energy
from the mechanical oscillator and offer at the same time an
extremely sensitive tool to read its mechanical fluctuations
\cite{PRLArcizet}. The passive cooling technique
notably\cite{ConstanzeNature,Arcizet,Gigan,Kippenberg,IvanAPL,Mavalvala},
also called self-cooling technique, is analogous to Doppler or
cavity cooling of atoms
\cite{Hansch,Vuletic,PRLRitsch,Rempe,OpticsLetters} and avoids
adding noise in the mechanical system. It is therefore expected to
reach in principle physical fundamental limits. This technique
relies on the intrinsic back-action of light on mechanical degrees
of freedom in opto-mechanical systems where photothermal pressure or
radiation pressure effects can arise \cite{Braginsky}. It has
already been used to cool different kind of mechanical resonators,
ranging from millimetric mirrors \cite{Arcizet,Gigan} to AFM
microlever mirrors \cite{ConstanzeNature}, toroid microcavities
\cite{Kippenberg} and wavelength sized micro-mirrors \cite{IvanAPL}.
In all these cases, the mechanical resonator to be cooled must
confine the light in an optical cavity and has therefore to be
larger than the wavelength of photons. In this paper, we describe a
novel method for cooling optically and passively the motion of a
nanomechanical resonator smaller than the wavelength of photons. Our
proposal, using back action in a situation of intense coupling
between optics and nanomechanics, breaks the diffraction barrier
limit allowing the investigation of quantum phenomena in
nanomechanical systems \cite{PRLImamoglu}.


The principle for achieving cooling is the following: the
nanomechanical resonator fluctuates under Brownian motion with a
noise distribution peak at its lowest eigenfrequency $f_{0}$. When
placed in the mode of a high finesse Fabry-Perot cavity at resonance
with a laser of wavelength $\lambda$, it scatters the cavity
photons, as depicted in figure 1. Because photons circulate several
times back and forth in the cavity, they need a finite time $\tau$
to reach a new equilibrium after each scattering event. This induces
a delay in their back action on the motion of the resonator trough
radiation pressure or dipolar forces. Exploiting this retarded
back-action, we show here that an additional optically induced
viscous damping is obtained, resulting in a net cooling of the
nanoresonator vibrational fluctuations. Photons leaving the cavity
carry away the excess energy.

Cavity cooling of atoms was already explored exploiting a detuning
of the cavity induced by the dispersive response of a atom moving in
the cavity \cite{PRLRitsch,Vuletic,Rempe}. Aiming here at cooling a
solid-state nanomechanical object, we strongly deviate from this
atomic case. Losses mechanisms typical for solid-state systems, such
as exciton or polariton absorption resonances, interband and
intraband absorption, or even Rayleigh scattering out of the cavity
call for a novel approach including both the absorptive and
dispersive optical response of the nanomechanical scatterer placed
in the cavity. The problem of a scattering element placed in a
focused gaussian beam in the paraxial limit has been shown to be
nearly equivalent to that of a thin plane of conductivity $\sigma$
at optical frequencies placed in a plane wave
\cite{JorgBirthday,ThesisHögele,VanEnk}. Therefore we model the
nanoscatterer as a thin plane of transmittance $1/(1+\Sigma)$ and
reflectance $-\Sigma/(1+\Sigma)$ where $\Sigma$ consists of a real
and of an imaginary part $\Sigma=\Sigma_{1}+i\Sigma_{2}$ and where
$\Sigma=\sigma/2\epsilon_{0}c$. The prescription to determine
$\Sigma_{1}$ and $\Sigma_{1}$ relies on a simple measurement of the
reflectance and the transmittance of the nanoscatterer in a focused
gaussian beam of size matching the cavity mode. The field amplitude
distribution in the cavity perturbed by the nanoscatterer are then
computed using a system of three transfer matrixes for plane waves
\cite{JorgBirthday}.

In an empty Fabry-Perot cavity with mirrors of high reflectance $r$,
a laser coupled resonantly creates a steady-state intensity
distribution at position $x_{0}$ along the optical axis that can be
approximated with $P(x_{0})=2gP_{0}sin^{2}(kx_{0})$ (figure 1),
where $P_{0}$ is the optical power impinging on the cavity,
$k=(2\pi/\lambda)$ is the wave number and g the cavity amplification
factor, which relates to the cavity finesse f such $g=(2f)/\pi\simeq
2/(1-\mid r\mid^{2})$. The nanoresonator is first placed at position
$x_{0}$ along this intensity distribution. The cavity is then tuned
to its maximum of transmission by ajusting the back mirror position,
leading to a transmission
$T(x_{0})=1/[1+\Sigma_{1}(1+2g\sin^{2}(kx_{0}))]^{2}$. The cavity is
thereafter actively stabilized on this resonance to enhance the
interaction of photons with the nanoresonator, as well as to reduce
the contribution of the laser noise on the noise of the transmitted
photons. At finite temperature, the scatterer fluctuates around
$x_{0}$ with a small amplitude $x$ such $kx\ll 1$, leading to a
fluctuation of the transmission $T(x_{0})$ which can be written
after a straightforward but cumbersome calculation:
\begin{equation}
T(x_{0},x)=1/[H_{1}(x_{0},x)+H_{2}(x_{0},x)]
\end{equation} with \begin{equation}
H_{1}(x_{0},x)=[1+\Sigma_{1}(1+2g\sin^{2}(kx_{0}))+2g\Sigma_{1}\sin(2kx_{0})kx]^{2}
\end{equation} and
$H_{2}(x_{0},x)=4g^{2}\Sigma_{2}^{2}\sin^{2}(kx_{0})(k^{2}x^{2})$.
The transmission fluctuation enables read-out of the nanomechanical
resonator motion linear in x, provided that $\Sigma_{1}$ is
non-zero. The purely dispersive effects relating to $\Sigma_{2}$
contribute only to a non-linear response in $x^{2}$. Loss
mechanisms, for instance escape of photons out of the cavity or
absorption by the nanoscatterer, are therefore mandatory for an
efficient read-out of the nanomechanical motion fluctuation. The
location along axis $x_{0}$ of maximum sensitivity for this read-out
(i.e. the extrema of dT/dx) depends on the factor $g\Sigma_{1}$,
which is the ratio between the nanoresonator-induced losses and the
intrinsic losses of the cavity.

The essence of passive cooling relies on a delayed response of the
back-action force F acting on the mechanical system, here the
nanoresonator. The effective temperature $T_{eff}$ reached by
cooling can be written $T_{eff}=T_{b}(\Gamma/\Gamma_{eff})$ where
$\Gamma_{eff}$ is the optically modified damping rate of the
nanomechanical resonator, $\Gamma$ being its natural damping and
$T_{b}$ the bath temperature. $\Gamma_{eff}$ is in a classical limit
given by \cite{ConstanzeNature}
$\Gamma_{eff}=\Gamma[1+Q_{m}(\omega_{0}\tau/(1+\omega_{0}^{2}\tau^{2}))\nabla
F/K]$, where the gradient of the force upon nanoresonator motion
$dF(x_{0},x)/dx$ is noted $\nabla F$, $Q_{m}=\omega_{0}/\Gamma$ is
the mechanical quality factor of the nanoresonator, $\omega_{0}=2\pi
f_{0}$ its eigenfrequency, K its spring constant and $\tau$ the
delay time of the force. In an empty cavity at resonance with
transmission $T_{0}$, the delay time of photon pressure on the
mirrors is given by the empty cavity storage time
$\tau_{c}=g\tau_{0}$ where $\tau_{0}$ is the time of flight of
photons trough the cavity. When the nanoresonator is placed at
position $x_{0}$ in the cavity and the cavity hold on resonance with
an average transmission $T(x_{0})$, the storage time is modified to
$\tau=\tau_{c}\sqrt{T(x_{0})/T_{0}}$. We will study cases where the
variation of the storage time over the nanoresonator position
fluctuations is negligible.


The optical force $F_{0}$ acting on the nanoresonator when placed at
the waist of a forward propagating gaussian wave of power $P_{0}$
matched to the geometry of the cavity mode can be obtained
calculating the transmission and reflection by the nanoscatterer and
relating it to a net momentum transfer
$F_{0}=(2\Sigma_{1}+2\Sigma^{2}_{2}-2\Sigma^{2}_{1})P_{0}/c$. At
first order, this force is proportional to $\Sigma_{1}$ and relies
then on loss mechanisms. In our cavity scheme, the nanoscatterer is
placed at $x_{0}$ and the cavity maintained at resonance. In this
configuration, a calculation of the energy-flow imbalance between
the waves travelling forward and backward on both sides of the
fluctuating nanoresonator leads to the static photon-induced force
$F(x_{0},x)$ acting on it: $F=(P_{0}/c)[G_{1}+G_{2}]T$ where
$G_{1}=2\Sigma_{1}[2g\Sigma_{1}\sin^{2}(k(x_{0}+x))+\Sigma_{1}\sqrt{2g}\sin(k(x_{0}+x))\cos(k(x_{0}+x))+(1+\Sigma_{1})]$
and
$G_{2}=4g\Sigma^{2}_{2}\sin^{2}(k(x_{0}+x))-2g\Sigma_{2}\sin(2k(x_{0}+x))-2\Sigma_{2}\sqrt{2g}\cos(2k(x_{0}+x))+2\sqrt{2g}\Sigma^{2}_{2}\sin(2k(x_{0}+x))+2\Sigma^{2}_{2}$.
We will focus on situations where scattering of photons by the
nanoresonator is small ($\Sigma_{1},\Sigma_{2}\ll1$) and the finesse
of the cavity large $g\gg1$.

We first study the case where dispersive scattering is very weak
($\Sigma_{2}\simeq0$) and losses induced by the presence of the
nanoresonator are dominating other optical losses in the cavity
$g\Sigma_{1}\gg1$. Figure 2 displays the case $g\Sigma_{1}=10$, with
$g=2\times10^{4}$ and $\Sigma_{1}=5\times10^{-4}$. In this so-called
lossy limit, we see in figure 2 that the force F always points along
the incoming photons, which is a reminder of the broken symmetry of
the system induced by the presence of the laser source on the left
hand-side of the cavity. The amplitude of the force is at most equal
to $F_{0}$, the force without cavity. The cavity does not amplify
the force but provides it with a gradient over x and a retardation.
As a first numerical illustration, we consider the case of a
cylindrical single wall carbon nanotube of radius $r = 0.8$ nm and
length $l = 5$ $\mu$m oscillating at $f_{0} = 205$ KHz according to
the formula $f_{0}=0.281(r/l^{2})\sqrt{E/\rho}$ \cite{Tischenko}
(Young modulus E$=1$ TPa and density $\rho=1.925$g/cm$^{3}$ for a
single wall nanotube) with a spring constant $K=7.7\times10^{-9}$N/m
according to $K=(3\pi/4)Er^{4}/l^{3}$, a mechanical quality factor
$Q_{m}= 10^{3}$ \cite{Zant} and $\Sigma_{1}=5\times10^{-4}$ placed
in a cavity of $g=2\times10^{4}$ \cite{collaboration} at position of
maximum (dF/dx) as indicated in figure 2. For a cavity of $50$
$\mu$m length illuminated with $P_{0}=1$ mW of laser power at
$\lambda=780$ nm, this would lead to $\Gamma_{eff}\simeq \Gamma(1\pm
47)$. When $\Gamma_{eff}<0$ \cite{ConstanzeSelfOscillation}, a
regime of mechanical self oscillation is reached and governed by a
purely lossy mechanism in the radiation pressure, in contrast to
recently developed optical back-action schemes for mechanical
resonators
\cite{ConstanzeNature,Arcizet,Gigan,Kippenberg,IvanAPL,Mavalvala}.
In a second numerical illustration, we consider a cylindrical
diamond nanorod (E$=1.14$ TPa and $\rho=3.52$g/cm$^{3}$) of radius
$r = 4$ nm, length $l = 0.5$ $\mu$m, $K=5\times10^{-3}$N/m
oscillating at $81$ MHz with $Q_{m}= 10^{4}$ \cite{Houston}. The
cavity-induced modification of the damping becomes
$\Gamma_{eff}/\Gamma\simeq(1\pm 7\times10^{-2})$, namely negligible.
Still, the interest of using such a stiff nanostructure will appear
in the following section.


\vspace{0.1cm}

We now study the contribution of dispersive effects related to
$\Sigma_{2}$. Figure 3 displays the optical force F acting on the
nanoresonator when $g=2\times10^{4}$ and
$\Sigma_{2}=5\times10^{-4}$. Using a non-vanishing
$\Sigma_{1}=10^{-5}\ll\Sigma_{2}$ allows for a read-out of x in the
transmission. The force can now be made positive or negative,
mimicking dipolar forces. Its direction depends on the energy
transfer between the two optical resonators formed on the left
hand-side of the cavity by the input mirror and the nanoscatterer
and on the right hand-side by the nanoscatterer and the back mirror.
Interestingly enough and in contrast to the purely lossy case, F is
here amplified by the cavity. The cavity role is to provide as
before a force gradient, a retardation, but on top of that an
amplification of the force. This three-fold advantage is apparent in
the following numerical illustration. We consider the same carbon
nanotube as previously and we find that the maximum damping
amplification factor becomes $\Gamma_{eff}/\Gamma=4.7\times10^{5}$,
four orders of magnitude larger than in the lossy case. At the same
time, an optical spring effect occurs, modifying the effective
spring constant to $K_{eff}=K[1-(1/(1+\omega_{0}^{2}\tau^{2}))\nabla
F/K]$ and preventing efficient cooling by driving the nanotube into
an instability regime already at moderate power \cite{LongPapier}.
The cooling is practically limited to a temperature $T_{eff}/T=0.2$.
In contrast, the diamond nanorod is stiff enough to preclude early
appearance of instabilities, allowing to reach under the same
experimental conditions a temperature reduction factor of $700$.
Starting from a standard liquid helium bath environment at 1.8 K,
the cooling mechanism would bring the motion of the diamond nanorod
to an effective temperature of $2.5$ mK, at which the quantum limit
is reached.

The study of small size nanoresonators, like carbon nanotubes or
diamond nanorods, appears then extremely appealing for investigation
of optomechanical phenomena at the nanoscale. However we should note
that reducing the size of the nanoresonator may also lead to a
reduction of $|\Sigma|$ and hence of the force F. The advantage of
using a solid state system is then apparent since it allows a wide
variation of $\Sigma_{1}$ and $\Sigma_{2}$, by choosing the photons
wavelength or the material of the nanoscatterer, and hence a direct
engineering of the cooling efficiency. Using an excitonic resonance
in a carbon nanotube \cite{Heinz} or an implanted nitrogen-vacancy
impurity resonance in diamond \cite{Wrachtrup} for instance would
allow to amplify both the absorptive and the dispersive response,
with a controllable ratio between the two. The high aspect ratio of
nanorods offers additionally the possibility of a selective coupling
to distinct polarisation modes of the cavity and hence a
supplementary degree of freedom in the optimization of cooling.
Together with the reduction of the number of phonon channels
eventually coupling the nanoresonator to the thermal bath, it makes
nanorod structures very promising candidates to study mechanical
quantum phenomenona at the nano-microscale.

\vspace{0.1cm}

In conclusion, we have studied a cavity cooling scheme of a
nanomechanical resonator scattering photons within a high finesse
Fabry-Perot cavity. This geometry, on top of allowing using
sub-wavelength sized objects, offers the possibility to engineer
separately ultra high finesse cavities on the one hand and high
quality factor mechanical resonators on the other hand, an issue
usually limiting opto-mechanical cooling experiments. Cavity cooling
in a purely dispersive limit and in a detuned cavity was discussed
for atoms and molecules \cite{PRLRitsch,Vuletic,Rempe} and very
recently for a macroscopic membrane \cite{Harris,Meystre}. Our
scheme relies here on a tuned cavity and in contrast to molecules or
atoms, the nanoresonator is attached to a holder and cannot escape
out of the cavity, offering the possibility to set the motion of the
nanoresonator into self-oscillation \cite{ConstanzeSelfOscillation}.
The onset of non linear opto-mechanical behaviors in this new
situation of intense coupling between nanomechanics and optics will
require a more complete theory, as developed in ref \cite{Marquardt}
to study multi-stabilities in Fabry-Perot cavities. Optical control
of vibration properties of nanomechanical systems would open new
routes, not only for sensing applications, but also for testing the
quantum mechanical description of tiny objects.

We thank C. Metzger, J. P. Kotthaus, J. Reichel, T. W. H\"{a}nsch,
F. Marquardt, T. Kippenberg and B. Lorenz for stimulating
discussions. I. Favero acknowledges support of the Alexander Von
Humboldt foundation.

\newpage

\begin{figure}[t]
\hspace{0.1cm}
\includegraphics[width=15cm]{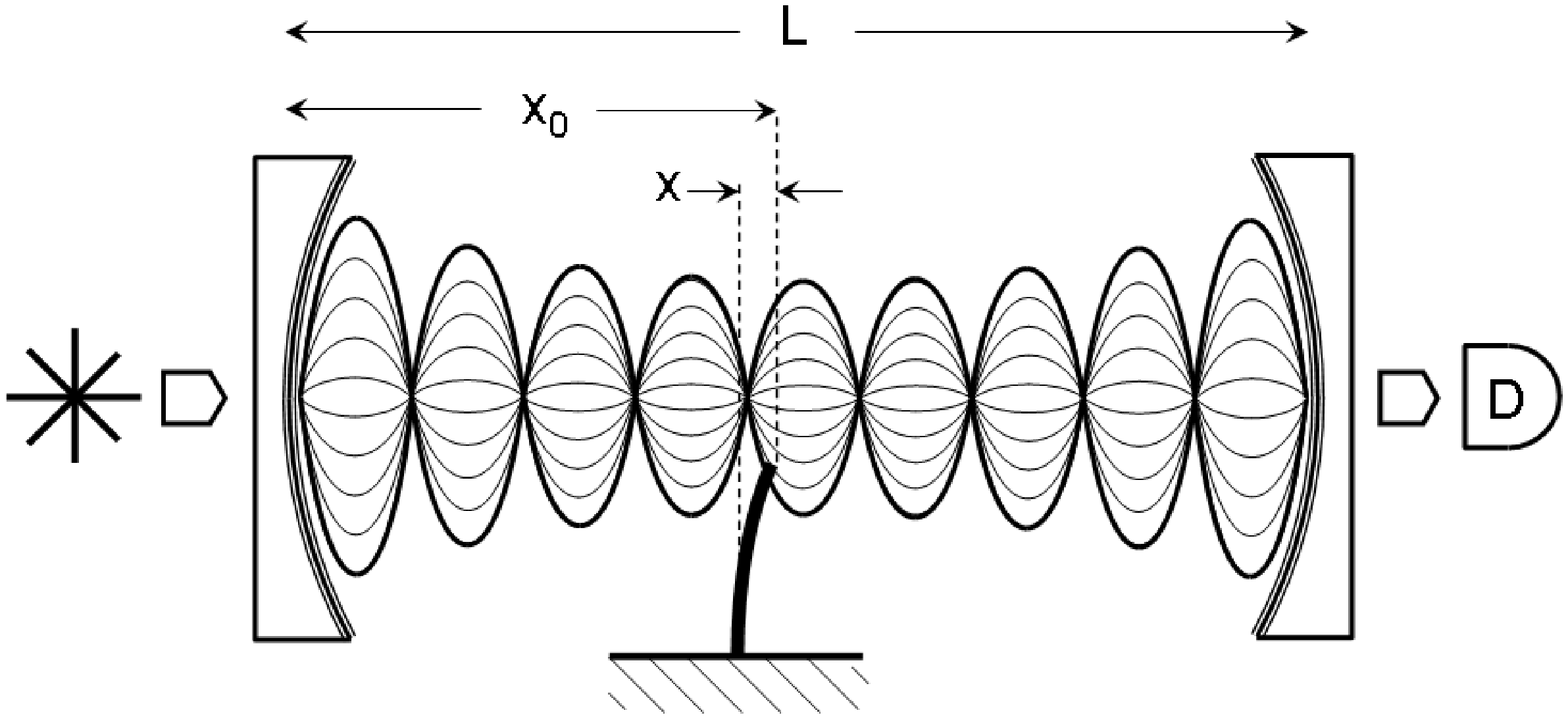} \hspace{+0.4cm} \caption{A nanomechanical resonator scattering
photons in a high finesse Fabry-Perot cavity resonantly coupled to a
laser on its left-hand side.} \label{resonance_combs}
\end{figure}

\begin{figure}[t]
\hspace{-0.1cm}
\includegraphics[width=15cm]{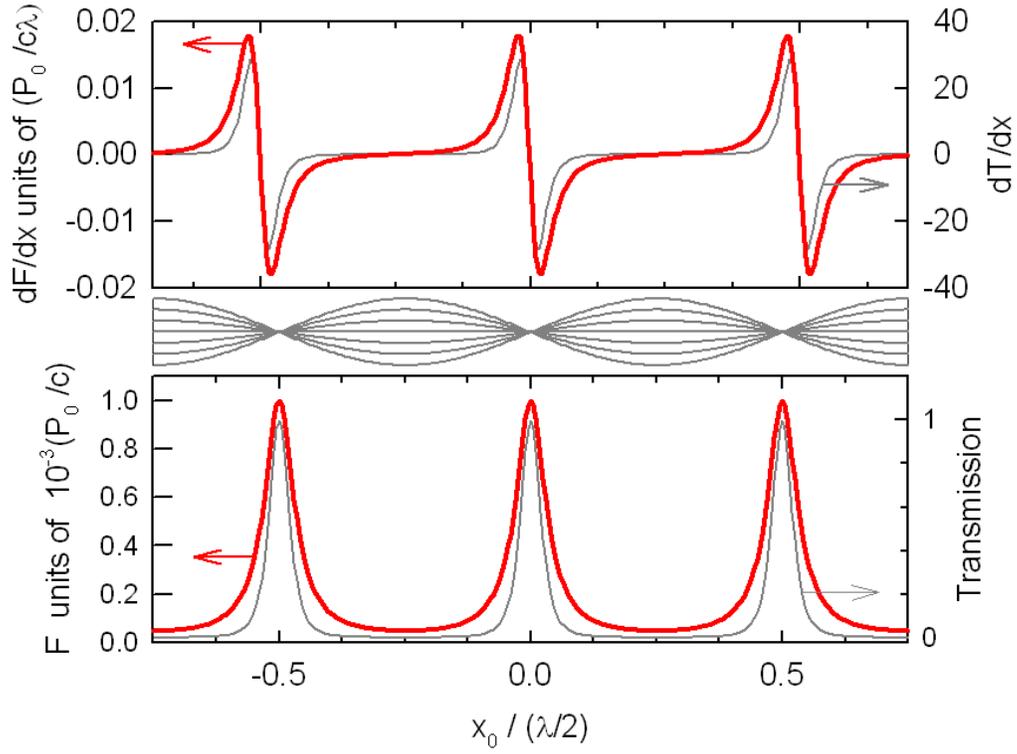} \hspace{+0.4cm} \caption{Case: $g\Sigma_{1}=10$,
with $g=2\times10^{4}$ and $\Sigma_{1}=5\times10^{-4}$. Bottom:
Transmission of the cavity $T(x_{0})$ and optical force F for zero
motion fluctuation of the nanoresonator ($x=0$). Middle: amplitude
snap-shots of the standing wave in the corresponding empty cavity.
Top: Gradient of the force and of the transmission upon the
nanoresonator fluctuation x, as a function of its average position
$x_{0}$.} \label{resonance_combs}
\end{figure}

\begin{figure}[t]
\hspace{-0.1cm}
\includegraphics[width=15cm]{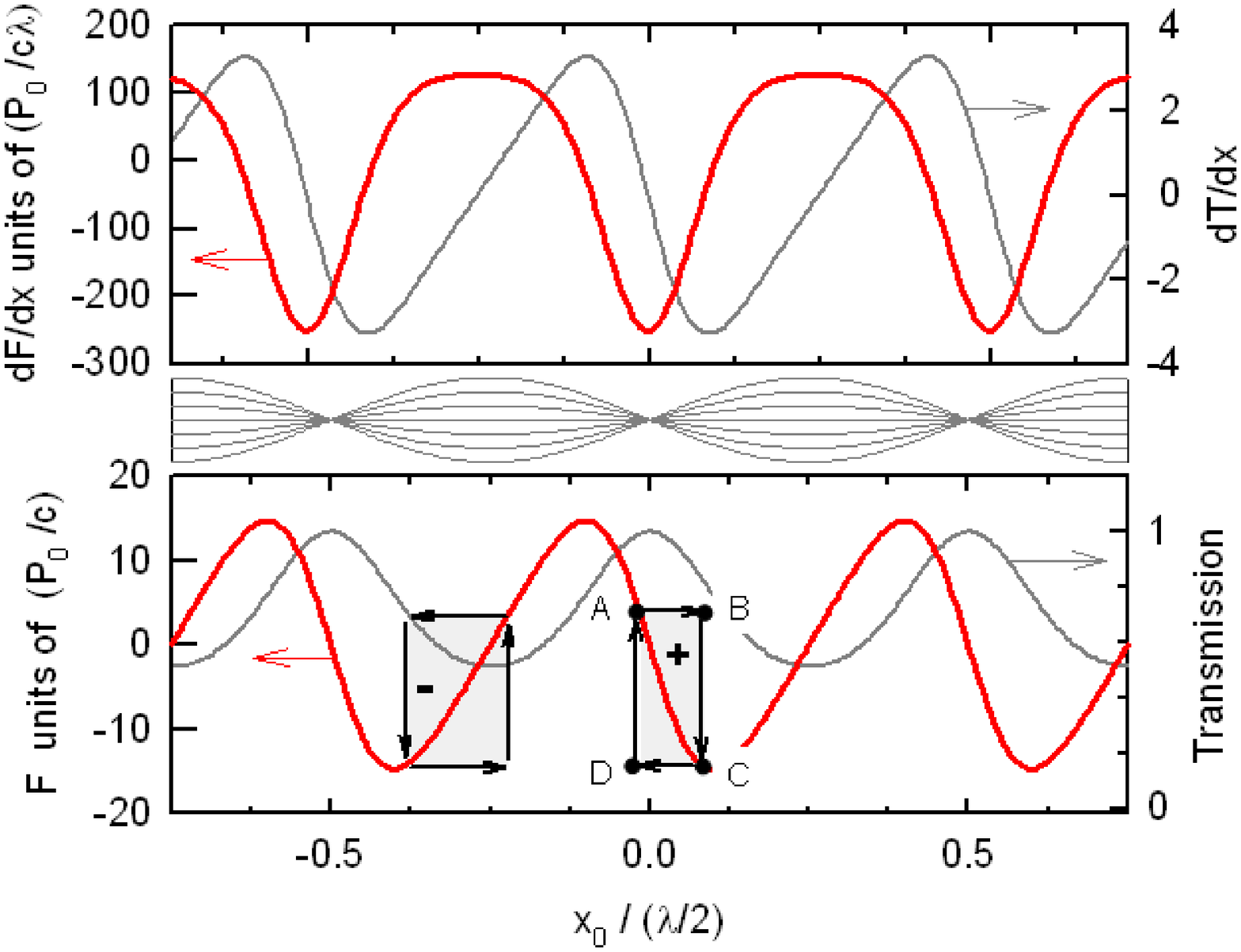} \hspace{+0.4cm} \caption{Case: $g=2\times10^{4}$,
$\Sigma_{1}=10^{-5}$ and $\Sigma_{2}=5\times10^{-4}$. The optical
modification of the nanoresonator damping can be described in a
thermodynamical manner following the cycles on the lower part of the
figure. The nanoscatterer moves suddenly from point A to point B,
leaving no time to the intensity distribution in the cavity and
hence the force to follow. Waiting long enough from B to C allows
the force to slowly recover its equilibrium level. Closing back the
cycle leads to a hysteresis which is characteristic of an
irreversible energy transfer and produces a net viscous force on the
nanoresonator, which can damp or amplify the motion
\cite{Braginsky,ConstanzeNature}.} \label{resonance_combs}
\end{figure}


\begin{references}

\bibitem{Legett}
A. J. Legett, J. Phys. Condens. Matter, {\bf 14}, R415-R451 (2002).

\bibitem{Haroche}
S. Haroche, Physics today, 36 (1998).

\bibitem{Bouwmeester}
W. Marshall, C. Simon, R. Penrose, and D. Bouwmeester, Phys. Rev.
Lett. {\bf 91}, 130401 (2003).

\bibitem{Penrose}
R. Penrose, in Mathematical Physics 2000 edited by A. Fokas, A.
Grigoryan, T. Kibble, and B. Zegarlinski (Imperial College, London).

\bibitem{Schwab}
M. D. LaHaye, O. Buu, B. Camarota, and K. C. Schwab, Science {\bf
304}, 74 (2004).

\bibitem{Cleland}
R. G. Knobel, A. N. Cleland, Nature {\bf 424}, 291 (2003).

\bibitem{Mancini}
S. Mancini, D. Vitali, and P. Tombesi, Phys. Rev. Lett. {\bf 80},
688 (1998).

\bibitem{Cohadon} P. F. Cohadon, A. Heidmann, and M. Pinard,
Phys. Rev. Lett. {\bf 83}, 3174 (1999).

\bibitem{BouwmeesterActive} D. Kleckner and D. Bouwmeester,
Nature {\bf 444}, 75 (2006).

\bibitem{ConstanzeNature} C. H\"{o}hberger Metzger and K. Karrai, Nature {\bf 432}, 1002 (2004).

\bibitem{Arcizet} O. Arcizet, P. F. Cohadon, T. Briant, M. Pinard and A. Heidmann,
  Nature {\bf 444}, 71 (2006).

\bibitem{Gigan} S. Gigan, H. R. B\"{o}hm, M. Paternostro, F. Blaser, G. Langer, J. B. Hertzberg, K. Schwab, D. B\"{a}uerle, M. Aspelmeyer and A. Zeilinger, Nature {\bf 444}, 67 (2006).

\bibitem{Kippenberg} A. Schliesser, P. Del'Haye, N. Nooshi, K. J. Vahala, and T. J. Kippenberg, Phys. Rev. Lett. {\bf 97}, 243905 (2006).

\bibitem{IvanAPL} I. Favero, C. Metzger, S. Camerer, D. K\"{o}nig, H. Lorenz, J. P. Kotthaus, and K.
Karrai, Appl. Phys. Lett. {\bf 90}, 104101 (2007).

\bibitem{Mavalvala} T. Corbitt, Y. Chen, E. Innerhofer, H. M\"{u}ller-Ebhardt, D. Ottaway, H. Rehbein, D. Sigg, S. Whitcomb, C. Wipf, and N. Mavalvala, Phys. Rev. Lett. {\bf 98}, 150802 (2007).

\bibitem{PRLArcizet} O. Arcizet, P. F. Cohadon, T. Briant, M. Pinard, A. Heidmann, J.-M. Mackowski, C. Michel, L. Pinard, O. Francais, and L. Rousseau, Phys. Rev. Lett. {\bf 97}, 133601 (2006).

\bibitem{Hansch} T. W. H\"{a}nsch, and A.L. Schawlow, Optics Commun. {\bf 13}, 68 (1975).

\bibitem{PRLRitsch}
P. Horak, G. Hechenblaikner, K. M. Gheri, H. Stecher, and H. Ritsch
, Phys. Rev. Lett. {\bf 79}, 4974 (1997).

\bibitem{Vuletic} V. Vuletic and S. Chu , Phys. Rev. Lett. {\bf 84}, 3787 (2000).

\bibitem{Rempe} P. Maunz, T. Puppe, I. Schuster, N. Syassen, P. W. H. Pinkse, and G.
Rempe, Nature {\bf 428}, 50 (2004).

\bibitem{OpticsLetters} K. Karrai, I. Favero, and C. Metzger, arXiv {\bf 0706.2841}(2007).

\bibitem{Braginsky} V. B. Braginsky and A. B. Manukin, {\it Measurements of weak forces in Physics experiments} (Chicago University Press, Chicago,1977).

\bibitem{PRLImamoglu}
I. Wilson-Rae, P. Zoller, and A. Imamoglu, Phys. Rev. Lett. {\bf
92}, 075507 (2004).


\bibitem{JorgBirthday}
K. Karrai, and R. J. Warburton, Superlattices and Microstructures
{\bf 33}, 311 (2003).

\bibitem{ThesisHögele} A. H\"{o}gele, PhD Thesis LMU (2006).

\bibitem{VanEnk}
S. J. van Enk, and H. J. Kimble, Phys. Rev. A {\bf 63}, 023809
(2001).

\bibitem{Tischenko} W. Weaver, S. P. Timoshenko, and D. H. Young, {\it Vibration problems in engineering}, John Wiley and Sons, (1990).

\bibitem{Zant} B. Witkamp, M.Poot, and H. S. J. van der Zant, Nano Letters {\bf 6}, 12, 2904 (2006).

\bibitem{collaboration} Y. Colombe, T. Steinmetz, G. Dubois, F. Linke, D. Hunger, and J. Reichel, Nature {\bf 450}, 272 (2007). We are currently using a similar type of
cavity. I. Favero, D. Hunger, S. Stapfner, J. Reichel, E. Weig, H.
Lorenz, and K. Karrai, unpublished.

\bibitem{ConstanzeSelfOscillation}
C. H\"{o}hberger Metzger and K. Karrai, Nanotechnology, 4th IEEE
Conference on Nanotechnology, 419-421 (2004).

\bibitem{Houston}
J.W. Baldwin, M.K. Zalalutdinov, T. Feygelson, B.B. Pate, J.E.
Butler, B.H. Houston, Diamond and related materials, {\bf 15}, 2061
(2006).

\bibitem{LongPapier}
C. Metzger, I. Favero, A. Ortlieb, and K. Karrai, arXiv {\bf
0707.4153}.

\bibitem{Heinz} F. Wang, G. Dukovic, L. E. Brus, and T. F. Heinz, Science {\bf 308}, 838 (2005).

\bibitem{Wrachtrup} J. Meijer, B. Burchard, M. Domhan, C. Wittmann, T. Gaebel, I. Popa, F. Jelezko, and J. Wrachtrup, Appl. Phys. Lett. {\bf 87}, 261909 (2005).

\bibitem{Harris} J. D. Thompson, B. M. Zwickl, A. M. Jayich, F. Marquardt, S. M. Girvin, and J. G. E. Harris, arXiv
{\bf 0707.1724}.

\bibitem{Meystre} M. Bhattacharya, H. Uys, and P. Meystre, arXiv {\bf 0708.4078}.

\bibitem{Marquardt} F. Marquardt, J. G. E. Harris, S. M. Girvin, Phys. Rev. Lett.
{\bf 96}, 103901 (2006).




\end{references}
\end{document}